%% file: main.tex
\documentclass[preprint]{elsarticle}
\input{defs}
\begin{document}
\begin{titlepage}
\pubblock
%\pubdate
%
\vfill
\def\thefootnote{\fnsymbol{footnote}}
\Title{\LARGE \sffamily \bfseries
How well can we guess theoretical uncertainties?
\support
}
\vfill
\Author{\normalsize \bfseries \sffamily
Andr\'{e} David \email{andre.david@cern.ch}}               
\Address{PH Department, CERN, Switzerland}
\bigskip
\Author{\normalsize \bfseries \sffamily
Giampiero Passarino \email{giampiero@to.infn.it}}               
\Address{\csumb}

\vfill
\vfill
\begin{Abstract}
\noindent 
The problem of estimating the effect of missing higher orders in perturbation theory
is analyzed with emphasis in the application to Higgs production in gluon-gluon fusion.
Well-known mathematical methods for an approximated completion of the perturbative series are applied with the goal to not truncate the series, but complete it in a well-defined way, so as to increase the accuracy - if not the precision - of theoretical predictions.
The uncertainty arising from the use of the completion procedure is discussed and a recipe for constructing a corresponding probability distribution function is proposed.
\end{Abstract}
\vfill
\begin{center}
%\KW{Feynman diagrams, loop calculations, radiative corrections, effective Lagrangian,
%Higgs physics} 
%%Keywords: Feynman diagrams, loop calculations, radiative corrections, effective Lagrangian,
%%Higgs physics \\[5mm]
%\PACS{11.15.Bt, 12.38.Bx, 13.85.Lg, 14.80.Bn, 14.80.Cp}
%%PACS classification: 11.15.Bt, 12.38.Bx, 13.85.Lg, 14.80.Bn, 14.80.Cp
\emph{Submitted to Phys. Lett. B}
\end{center}
\end{titlepage}
\def\thefootnote{\arabic{footnote}}
\setcounter{footnote}{0}
%--
\small
\thispagestyle{empty}
\tableofcontents
\normalsize
%--
\clearpage
%--
\setcounter{page}{1}

%--
\section{Introduction \label{Intro}}
%--
In the past $30$ years, the commonly accepted way to estimate theoretical uncertainties associated to collider physics observables has been based on the notion of QCD scale variations.
We introduce the concept of {\bf{MHO}}({\bf{U}}), missing higher order (uncertainty), which is linked to the 
truncation error in the perturbative expansion.
At present, and for some time to come, estimations of observables will be based on a finite number of terms of a series, such that additional information on the behavior of that series should be exploited.

Regardless of their precision, truncated calculations are only as accurate as the higher orders that they lack.
A more accurate evaluation of the observable may be obtained by estimating the MHO.
The issue of precision then becomes more tightly bound to the estimation of the MHOU, taking into account both the uncertainty on the MHO estimation procedure as well as any uncertainties in the terms that have already been calculated.

In this Letter, the problem of MHO(U) in Higgs production through gluon-gluon fusion is approached using sequence transformations to improve the rate of convergence of the series and directly estimate the MHO.
In \Sref{sec:beyond} we discuss foundational issues related to the MHO problem and the applicability of sequence transformations.
In \Sref{sec:ggFinputs} we summarize existing calculations of Higgs production through gluon-gluon fusion, mapping out the inputs needed to estimate the MHO.
Then, in \Sref{sec:seqtrans}, we introduce different types of sequence transformations and discuss in detail their performance in synthetic problems as well as applications to series involving physical observables.
The main results are then presented in \Sref{sec:ggF} where we apply sequence transformations to the problem of Higgs production through gluon-gluon fusion problem and propose an estimate of MHOU and its probability distribution function (pdf).
Finally, in \Sref{sec:conclusions} we summarize the main arguments and results.
%--
%\include{foundation}
%\include{ggFinputs}
%\include{seqtrans}
%\include{ggF}
%
\input{foundation}
\input{ggFinputs}
\input{seqtrans}
\input{ggF}
%%%%%\include{param}
%--
%\clearpage
%--
\section{Conclusions \label{sec:conclusions}}
%--
The flat part of the MHOU pdf has been chosen observing that $\sigma^{\ssS,3}$ is the last
known term of the series, that known and predicted coefficients are all positive, and that all 
transforms ``predict'' convergence towards a value inside the interval of \eqn{bg} and close
to $\sigma^{\delta,5}$.
Therefore, our best guess is the one in \eqn{bg} since it would be ambitious to claim that $\sigma^{\tau,6}$ or $\sigma^{\delta,5}$ are the result, with a very 
small error.
One should mention in this regard that there is no proof of the uniqueness of the result reconstructed from its asymptotic series.
There is only evidence that all sequence transforms produce a result within a given interval to which we assign an uninformative prior, in the Bayesian sense.

It should be mentioned that we have included only the $\Pg\Pg\,$-channel.
At higher orders we have new channels, new color structures, etc.
For instance, the $\PQq\Pg\,$-channel contribution is negative; at low orders its contribution is sub-leading but nothing is known at higher orders.
This is a general problem that will affect all procedures aimed at estimating MHO(U).
Finally, it should be stressed that all re-summation procedures for non-alternating (divergent) series usually fail when the parameter expansion is on a (expected) cut in the complex plane.

We support the strategy presented for deriving information on MHO(U) with the following arguments:
%--
\bei

\item Given the (few) known coefficients in the perturbative expansion, we estimate the next (few) coefficients and the corresponding partial sums by means of sequence transformations. This is the first step towards ``reconstructing'' the physical observable in \eqn{asy}.

\item The use of sequence transformations was tried on a number of test sequences, including several physical observables.

\item A function can be uniquely determined by its asymptotic expansion if certain conditions are satisfied~\cite{Sokal:1980ey}.

\item The Borel procedure is a summation method which, under the above conditions, determines uniquely the sum of the series.
It should be taken into account that there is a large class of series that have Borel sums (analytic in the cut-plane) and there is evidence that Levin-Weniger transforms produce approximations to these Borel sums.
This is one of the plausibility arguments supporting our results.

\item The QCD scale variation uncertainty decreases when we include new (estimated) partial sums.

\item All known and predicted coefficients are positive and all transforms predict convergence within a narrow interval.

\item Missing a formal proof of uniqueness, we assume an uninformative prior between the last known partial sum and the (largest) predicted partial sum.

\eei

The arguments developed in this work support the opinion that perturbation theory up to N${}^3$LO is essential to obtain accurate definition of the theory (MHO) and shed some light on how to formulate consistent procedures for accurate computations (MHOU).
We conclude by saying that ``new'' insights into the properties of perturbative expansions are always important, since computing higher-order corrections is not only cumbersome and costly but also suffers fundamentally from the divergence of the series.

The investigation of QCD-scale and renormalization-scheme dependence of a truncated series should not be confused with the attempt to estimate its uncalculated remainder which is the true source of MHOU.

%--
%\clearpage

\section*{Acknowledgments}
We acknowledge the LHC Higgs Cross Section Working Group where the problem addressed here was posed and as the forum that brought the authors together.
AD recognizes useful discussions with Kirill~Melnikov at the XXV Rencontres de Blois.
GP acknowledges important discussions with Matteo~Cacciari and Stefano~Forte.

%--
\bibliographystyle{elsarticle-num}
\bibliography{main}{}

%\appendix
%\section{MHOU implementation}

%===
\end{document}

%% file: foundation.tex
\section{MHOU beyond scale uncertainties\label{sec:beyond}}

%--
Consider an observable $X\lpar Q,\mu\rpar$, where $Q$ is the typical scale of the process, and $\mu \equiv \{\muR,\muF\}$ are the renormalization and factorization scales.
The traditional procedure to estimate MHOU through scale variations~\cite{Dittmaier:2011ti} defines
%--
\bq
X^-_{\xi}\lpar Q,\mu\rpar = \min\Bigl\{
X\lpar Q\,,\,\frac{\mu}{\xi}\rpar\,,\,X\lpar Q\,,\,\xi\,\mu\rpar\Bigr\},
\quad
X^+_{\xi}\lpar Q,\mu\rpar = \max\Bigl\{
X\lpar Q\,,\,\frac{\mu}{\xi}\rpar\,,\,X\lpar Q\,,\,\xi\,\mu\rpar\Bigr\},
\eq
%--
or variations thereof, see \Bref{Cacciari:2011ze}.
Selecting a value for $\xi$ (typically $\xi = 2$) the prediction is that
%--
\bq
X^-_{\xi}\lpar Q,\mu\rpar < X\lpar Q,\mu\rpar < X^+_{\xi}\lpar Q,\mu\rpar.
\eq
%--
There are several examples in the literature where the $\xi = 2$ scale uncertainty of the $n$th order underestimates the $n+1$th order calculation.

There is also an open and debatable question on how to assign a probability distribution function (pdf) to the MHOU thus obtained \cite{THUTFindico}.
The procedure that is most commonly used is based on a Gaussian (or log-normal) distribution centered at $\mu = X_\cent = X\lpar Q,Q\rpar$.
This choice of central value is afflicted by the accuracy issues from truncation and there are cases in which the scale has been adapted to match resummation~\cite{mhovertwo,softgluonresummation}.
What to use for the standard deviation remains an open problem, though a common ansatz is to use $\sigma= \max\left(X^+_{2}(Q,Q), X^-_{2}(Q,Q)\right)$.
Alternatively, it could be assumed that the pdf is a uniform distribution
%--
\[
P(X) = \left\{
\begin{array}{ll}
\frac{1}{X^+_{2}(Q,Q) - X^-_{2}(Q,Q)} & X^-_{2}(Q,Q) < X < X^+_{2}(Q,Q), \\
0                    & \mbox{otherwise}.
\end{array}
\right.
.
\]
%--
Recently, Cacciari and Houdeau made a proposal to derive the pdf based on a flat (uninformative) Bayesian prior for the MHOU from the scale-variation prescription \cite{Cacciari:2011ze}.

More generally, the dependence on scales is only one part of the problem, as the MHO problem is based on how to interpret the relation between an observable ${\cal O}$, and a perturbative series
%--
\bq
{\cal O} \sim \sum_{n=0}^{\infty}\,c_n\,g^n.
\label{asy}
\eq
%--

The perturbative expansion of \eqn{asy} is unlikely to converge~\cite{Simon:1972yi} (see also \Brefs{Bender:1971gu,Bender:1990pd,Lipatov:1977hj,Buchvostov:1977fi,Brezin:1976wa,Brezin:1976vw}) and the asymptotic behavior of the coefficients is expected to be
%--
$
c_n \sim K\,n^{\alpha}\:n\,!/S^n
$
%--
when $n \to \infty$, and where $K, \alpha$ and $S$ are constants~\cite{Vainshtein:1994qq}.
An overview of the mathematical theory of divergent series and interpretation of perturbation series is given in \Bref{Suslov:2005zi}.

The requirement of \eqn{asy} ($\sim$) is not a formal one; it has the physical meaning of a smooth transition between the system with interaction and the system without it \cite{Fischer:1995cx}.
Furthermore, Borel and Carleman proved that there are analytic functions corresponding to arbitrary asymptotic power series~\cite{Hardy}. 
For a discussion on Borel summability and renormalon effects, we refer to the work of \Bref{Zakharov:1992bx}; for a criterion on Borel summability, we refer to the work of~\Bref{Sokal:1980ey}.

We also would like to mention that a procedure allowing for the elimination of the leading uncertainty of perturbative expansions in QCD can be found in \Bref{Beneke:1992ea} and that large orders in perturbation theory have been discussed in \Bref{Collins:1977dw}.

We should stress that recoverability of a function by means of its asymptotic series is possible only if ``enough'' analyticity is available~\cite{Fischer:1995cx} and any work on MHO(U) should be based on this assumption.
In other words, there are in general infinitely many functions with the same asymptotic expansion.
Therefore, one should assume that: a) there is a sufficiently large analyticity domain and b) that there is an upper bound on the remainder for each order above a certain value.
We will discuss the plausibility of these assumptions in the context of the example of Higgs production via gluon-gluon--fusion.
It is worth nothing that the authors of \Bref{Cacciari:2011ze} only assume b); starting from \eqn{asy}, they estimate the remainder $R_k = {\cal O} - \sum_{n=0}^{k}\,c_n\,g^n$, to be 
$R_k \approx c_{k+1}\,g^{k+1}$ with $c_{k+1} = \max\{\mid c_0\mid,\,\dots\,,\mid c_k\mid\}$. 
This, in turn, reflects into a width of $c_{k+1}\,g^{k+1}$ for the flat part of the uncertainty 
pdf.

Therefore, the MHO problem and its associated uncertainty can be summarized in one point: how can we make predictions for higher order perturbative coefficients, whose explicit calculation is cumbersome and time-consuming, while keeping a balance with analyticity?
As discussed in \Bref{Fischer:1995cx}, the problem is not that of divergence of the series, but of whether the expansion uniquely determines the function or not, and examples are given of functions which are singular at the origin while their asymptotic expansion is a convergent series.

We will not be able to answer these general questions (namely to prove uniqueness) and will rather concentrate on predicting higher orders using the well-known concept of ``series acceleration''~\cite{CVZ,BZ,BF}, i.e., one of a collection of sequence transformations for improving the rate of convergence of a series.
If the original series is divergent, the sequence transformation acts as an extrapolation method.
In the case of infinite sums that formally diverge, the helpful property of sequence transformations is that they may return a result that can be interpreted as the evaluation of the analytic extension of the series for the sum. 
The relation between Borel summation (the usual method applied for summing divergent series) and these extrapolation methods was noted for the first time in \Brefs{SB,SBB}.
Note that the definition of the sum of a factorially divergent series, including those with non-alternating coefficients, is always equivalent to Borel's definition (see Section~7 of \Bref{Suslov:2005zi}).

%% file: ggFinputs.tex
%--
\section{Existing calculations of Higgs production via gluon-gluon fusion}
\label{sec:ggFinputs}
%--
Let us consider what is presently known of Higgs production via gluon-gluon--fusion, \ie, the process $\Pg\Pg \to \PH$.
There have been several attempts to compute approximate N${}^3$LO corrections, see
\Brefs{Moch:2005ky,Ball:2013bra,Buehler:2013fha}.
Here we follow the work of \Bref{Ball:2013bra} and define
%--
\bq
\frac{\sigma_{\Pg\Pg}\lpar \tau\,,\,M^2_{\PH}\rpar }{ \sigma^0_{\Pg\Pg}\lpar \tau\,,\,M^2_{\PH}\rpar} =
K_{\Pg\Pg}\lpar \tau\,,\,M^2_{\PH}\,,\,\alphas\rpar
=
1 + \sum_{n=1}^{\infty}\,\alphas^n(\muR)\,K^n_{\Pg\Pg}\lpar \tau,\mu = \mh\rpar,
\label{FS}
\eq
%--
where $\tau = M^2_{\PH}/s$, $\sigma^0_{\Pg\Pg}$ is the LO cross section,
and the $K\,$-factor $K_{\Pg\Pg}$ was expanded in powers of $\alphas(\muR)$.
In \eqn{FS} it is understood that when computing the partial sums
%--
$
S_{\ssN} = 1 + \sum_{n=1}^{\ssN}\,\alphas^n(\muR)\,K^n_{\Pg\Pg}
$,
%--
$\alphas$ is computed at the highest level, \ie, NLO for $S_1$, NNLO for $S_2$, etc.

Introducing $\ugm_n = K^n_{\Pg\Pg}$, the known values are $\ugm_1=K^1_{\Pg\Pg}\lpar \tau,\mu = \mh\rpar = 11.879$ and $\ugm_2=K^2_{\Pg\Pg}\lpar \tau,\mu = \mh\rpar = 72.254$.
In their recent work, the authors of \Bref{Ball:2013bra} computed an approximation of $\alphas^3\lpar\mu\rpar\,
K^3_{\Pg\Pg}\,\lpar \mu\rpar$ at $\sqrt{s} = 8 \UTeV$ for $\mu=\mh/2$, $\mh$, and $2\mh$.
Since $\alphas^3\lpar\mu\rpar\, K^3_{\Pg\Pg}\,\lpar \mu\rpar$ is only known within a given interval (see Tab.~1 and discussion after Eq.(4.1) of \Bref{Ball:2013bra}) we report in \Tref{tab:ggFinputs} the numerical values of $\ugm_3$ as a central value ($\ugm^{\cent}_3$) and the corresponding uncertainty range ($\Delta\ugm_3$).
%--
%\bqa
%\ugm^{\cent}_3\lpar \mh/2\rpar \pm \Delta\ugm_3\lpar \mh/2\rpar 
%&=& 168.98 \pm 30.87,
%\nl
%\ugm^{\cent}_3\lpar \mh\rpar \pm \Delta\ugm_3\lpar \mh\rpar &=& 377.20 \pm 30.78,
%\nl
%\ugm^{\cent}_3\lpar 2\,\mh\rpar \pm \Delta\ugm_3\lpar 2\,\mh\rpar &=& 681.72 \pm 29.93.
%\eqa

\begin{table}
\begin{center}
\begin{tabular}{cc|ccc}
\hline
Our notation  & \Bref{Ball:2013bra} & $\mu=\mh/2$ & $\mu=\mh$ & $\mu=2\mh$ \\
\hline
 $\ugm_1$ & $K^1_{\Pg\Pg}$ & & 11.879 & \\ 
 $\ugm_2$ & $K^2_{\Pg\Pg}$ & & 72.254 & \\
 $\ugm^{\cent}_3 \pm \Delta\ugm_3$ & $K^3_{\Pg\Pg}$ & $168.98 \pm 30.87$ & $377.20 \pm 30.78$ & $681.72 \pm 29.93$\\ 
 \hline
\end{tabular}
\label{tab:ggFinputs}
\caption{Numerical values as derived from \Bref{Ball:2013bra} assuming $\sqrt{s} = 8 \UTeV$. These values are the relevant inputs to an estimation of MHO(U): while traditionally MHOU is estimated from the scale variation of $\ugm^{\cent}_3$, the proposed procedure only requires the values in the middle column ($\mu=\mh$).}
\end{center}
\end{table}

In \Tref{tab:ggFinputs} one can immediately see that the approximate calculation of $K^3_{\Pg\Pg}$ can be varied in two ways: 1) the change in $\ugm^{\cent}_3$ via scale variation, and 2) the intrinsic uncertainty $\Delta\ugm_3$ due to the approximate nature of the result.
While the traditional approach to MHOU estimation considers the effect from scale variation, the procedure that we put forth in later Sections combines $\Delta\ugm_3$ with the uncertainty on the estimation of the MHO based on sequence transformations.

%% file: seqtrans.tex
%--
\section{Sequence transformations}
\label{sec:seqtrans}
%--
The theory of sequence transformations is a well-established branch of numerical mathematics with many applications in science, as described in \Brefs{Weniger:1997zz,Jentschura:1999vn,Weniger:2001ct} and \Bref{Roy:2006cb}.
As an example in connection with the summation of the divergent perturbation expansion of the hydrogen atom in an external magnetic field, the work of \Brefs{Zama,Zamastil:2013jz} introduces a new sequence transformation which uses as input not only the elements of a sequence of partial sums, but also explicit estimates for the truncation errors.

Through sequence transformations, slowly convergent and divergent sequences and series can be transformed into sequences and series with hopefully better numerical properties.
Thus they are useful for improving convergence.
In most situations, a sequence transform does not sum a series exactly; however, in many cases, it correctly predicts some of the unknown terms of the sequence.

\subsection{The Levin $\tau\,$-transform}

Let us recall the definition of the Levin $\tau\,$-transform, see \Brefs{Lev,Wen,Wen2}.
Given the partial sums
%--
$
S_n = \sum_{i=0}^n\,\ugm_i\,z^i
$
%--
we define the $\tau\,$-transform as
%--
\bq
\tau^n_k(\beta) = \frac{\sum_{i=i_0}^k\,W^{\tau}\lpar n,k,i,\beta\rpar\,S_{n+i}}{\sum_{i=i_0}^k\,
W^{\tau}\lpar n,k,i,\beta\rpar},
\qquad \tau_k \equiv \tau^{0}_k \equiv \tau^0_k(0)
\label{Levintau}
\eq
where $i_0 = \max\{0,n-1\}$ and
$%\[
W^{\tau}(n,k,i,\beta) = (-1)^i\,
\binom{k}{i}
\frac{\lpar \beta+n+i\rpar_{k-1}}{\Delta S_{n+i-1}},
$%\]
%--
where $(z)_a = \Gamma(z+a)/\Gamma(z)$ is the Pochhammer symbol, and $\Delta$ is the usual forward-difference operator, $\Delta S_n = S_{n+1} - S_n$.
%-

The algorithm for estimating the first unknown coefficient is based on the Taylor expansion of $\tau_k$; if $S_1\,,\,\dots\,,\,S_k$ are known, one computes
%--
$
\tau_k - S_k = {\ougm}_{k+1}\,z^{k+1} + {\cal O}\lpar z^{k+2}\rpar
$
%--
and ${\ougm}_{k+1}$ is the prediction for $\ugm_{k+1}$.
%-
Of course, this prediction is not expected to be very reliable for small values of $k$.
Nevertheless, applying
%--
$
\tau_2 - S_2 = (\ugm^2_2/\ugm_1)\,z^3 + {\cal O}\lpar z^4\rpar
$
%--
to the series of \eqn{FS}, one predicts 
$
{\ougm}_3\lpar \mu = \mh\rpar = 439.48
$
which has the correct sign and the right order of magnitude when compared with the results from \Bref{Ball:2013bra},
$
346.42 < \ugm_3\lpar \mu = \mh\rpar < 407.48
$.
%--
\paragraph{Recursive estimation of unknown coefficients}
\label{sec:algo}

Let us outline our algorithm to improve the convergence of a series.
This algorithm can be used with any of the transforms introduced later and to any of the series also discussed in the examples later.
We give it below in an explicit form for the Levin $\tau\,$-transform $\tau^0_k(\beta)$ and applied to the series of \eqn{FS} assuming that the inputs from \Tref{tab:ggFinputs} are known:
%--
\begin{enumerate}

\item Use the first $3$ terms in \eqn{FS}, choose $\ugm_3= \ugm^{\cent}_3(\mu =\mh)$, and derive 
%--
$%\bq
\ougm_4 = 3\,\frac{\ugm_3}{\ugm_1 \ugm_2}\Bigl[
 2\,\frac{\lpar 5 + 2\,\beta\rpar\,\ugm^2_2 - \lpar 3 + \beta\rpar\,\ugm_1\,\ugm_3}
{12 + 7\,\beta + \beta^2}
 + \ugm_1\,\ugm_3 - \ugm^2_2 \Bigr].
%\label{bgamma}
$%\eq
%--
\item Construct $S_4$ assuming $\ugm_4 = {\ougm}_4$.
%--
\item Derive 
%--
$%\bq
\ougm_5 = 
\frac{\vartheta\,\ougm_4}{\ugm_1\,\ugm_2\,\ugm_3}\,
\lpar 120 + 72\,\beta + 15\,\beta^2 + \beta^3\rpar^{-1},
$%\eq
%--
%\bqa
%\vartheta &=& 240\,\ugm_1\,\ugm_2\,\ougm_4\,
%\lpar 1 + \frac{47}{6}\,\beta + \frac{1}{5}\,\beta^2 + \frac{1}{60}\,\beta^3\rpar
%          - 144\,\ugm_1\,\ugm^2_3\,
%\lpar 1 + \frac{13}{12}\,\beta + \frac{3}{8}\,\,\beta^2 + \frac{1}{24}\,\beta^3\rpar
%\nl
%{}&+&       24\,\ugm^2_2\,\ugm_3\,
%\lpar 1 + \frac{11}{6}\,\beta + \beta^2 + \frac{1}{6}\,\beta^3\rpar.
%\eqa
%--
where
$%\bq
\vartheta = 4\,\ugm_2^2\,\ugm_3 \, \Bigl(6 + 11\,\beta + 6\,\beta^2 + \beta^3\Bigr)
     - 6\,\ugm_1\,\ugm_3^2 \, \Bigl(24 + 26\,\beta + 9\,\beta^2 + \beta^3\Bigr)
     + 4\,\ugm_1\,\ugm_2\,\ougm_4 \, \Bigl(60 + 470\,\beta + 12\,\beta^2 + \beta^3\Bigr).
$%\eq
%--
\item Construct $S_5$ assuming $\ugm_5 = {\ougm}_5$.
%--
\item Repeat the previous steps until $\tau_3,\,\dots,\tau_6$ are constructed.
%--
\item Compare the $S_3,\,\dots,S_6$ with the $\tau_3,\,\dots,\tau_6$.
%--
\item Repeat steps 1--6 for $\ugm_3 = \ugm^{\cent}_3 +
\Delta\ugm_3$ and $\ugm_3 = \ugm^{\cent}_3 -
\Delta\ugm_3$, always taken at $\mu=\mh$.

\end{enumerate}
%--

The whole strategy is based on the fact that one can predict the coefficients by constructing an approximant with the known terms of the series ($\ugm_0,\dots,\ugm_n$) and expanding the approximant in a Taylor series.
The first $n$ terms of this series will exactly agree with those of the original series, while the subsequent terms may be treated as predicted coefficients.
Once the series is completed via an algorithm such as the one above, the dependence on $\mu$ is removed, and the notion of scale variation with it.
This implies that the uncertainty estimation is moved from scale variations to the completion procedure, as discussed in \Sref{sec:discussionMHOU}.
This procedure represents an extension of the work in \Bref{Cacciari:2011ze}.
%--
\paragraph{$\beta\,$-tuning}
\label{sec:betatuning}

If $\ugm_1,\dots,\ugm_3$ are known, the values of $\ugm_{1}$ and $\ugm_{2}$ can then be used to compute
%--
$
\beta= \lpar 1 - \frac{1}{2}\,\ugm_3\,\frac{\ugm_1}{\ugm^2_2}\rpar^{-1} - 2.
%\label{betat}
$
%--
This value of $\beta$ is such that $\ougm_3 = \ugm_3$.
With $\beta$ determined this way, one can then apply the recursive algorithm above.

For a discussion on $\beta\,$-tuning of the Levin $\tau\,$-transform, with applications to predicting new coefficients in the $g-2$ of muon and electron, see \Bref{Roy:2006cb}.
%For instance, given
Here, given
%--
$
a^{\PGm} = a\,\lpar 1/2 + 0.7655\,a + 24.05\,a^2 + 125.6\,a^3\rpar$,
where $a= \frac{\alpha}{\pi}
$
%--
and tuning $\beta = -0.90935$, one derives $\ougm_4 = 513.3$ with $\ugm_4$ expected in the range $433 < \ugm_4 < 713$.
%--

The $\beta\,$-tuned procedure is used to cross-check results without $\beta\,$-tuning in \Sref{sec:discussionMHOU}.
%--
\subsection{The Weniger $\delta\,$-transform}
A second transform that we have considered in detail in later Sections is the $\delta\,$-transform introduced by 
Weniger~\cite{Wen}:
%--
\bq
\delta_k(\beta) = \frac{\sum_{i=0}^k\,W^{\delta}\lpar k,i,\beta\rpar\,S_{i}}{\sum_{i=0}^k\,
W^{\delta}\lpar k,i,\beta\rpar},
\qquad \delta_k \equiv \delta_k(1),
\label{Wenigerdelta}
\eq
%--
where $%\[
W^{\delta}(k,i,\beta) = (-1)^i\,
\binom{k}{i}
\frac{\lpar \beta+i\rpar_{k-1}}{\lpar \beta+k\rpar_{k-1}}\,\frac{1}{\ugm_{i+1}\,z^{i+1}}.
$%\]
%--
Following the same algorithm described in \Sref{sec:algo}, the predicted $\ougm_n$ values for $\delta_k(1)$ are
%--
$%\bq
\ougm_4= 
   \frac{\ugm_3}{3\,\ugm_1\,\ugm_2}\,\lpar 4\,\ugm_1\,\ugm_3 - \ugm^2_2\rpar
$ and $
\ougm_5 = 
 \frac{\ougm_4}{10\,\ugm_1\,\ugm_2\,\ugm_3}\,\lpar
 \ugm^2_2\,\ugm_3 - 9\,\ugm_1\,\ugm^2_3 + 18\,\ugm_1\,\ugm_2\,\ougm_4 \rpar.
$%\eq
%--
\subsection{Other series transformations}
\label{sec:otherseqtrans}
There are other well-known transforms that we have tested using the algorithm described above:
%--
\bei
%--
\item Wynn's $\ep\,$-algorithm~\cite{Wynn}, the nonlinear recursive scheme
%--
$
\ep^n_{-1} = 0$, $\ep^n_0= S_n$,
$ 
\ep^n_{k+1} = \ep^{n+1}_{k-1} + \frac{1}{\ep^{n+1}_k - \ep^n_k}.
$
%--
\item Brezinski's $J\,$-algorithm~\cite{Brez}, based on the recursive scheme
%--
\bq
J^n_0= S_n, \quad
J^n_{k+1} = J^{n+1}_k - \frac{\Delta J^n_k\,\Delta J^{n+1}_k\,\Delta^2 J^{n+1}_k}
{\Delta J^{n+2}_k\,\Delta^2 J^n_k - \Delta J^n_k\,\Delta^2 J^{n+1}_k},
\quad
\Delta J^n_k= J^n_{k+1} - J^n_k.
\label{bre}
\eq
%--
\item Generalized Levin $t\,$-transform~\cite{Lev}:
%%--
%\bq
%t^n_k = \frac{\sum_{i=i_0}^k\,W^t\lpar n,k,i\rpar\,S_{n+i}}{\sum_{i=i_0}^k\,W^t\lpar n,k,i\rpar},
%\eq
%%--
%\[
%W^t(n,k,i) = (-1)^i\,
%\left(
%\begin{array}{c}
%k \\
%i
%\end{array}
%\right)
%\frac{\lpar n+i\rpar_{k-1}}{\Delta S_{n+i-1}}.
%\]
%
$
t^n_k = \frac{\sum_{i=i_0}^k\,W^t\lpar n,k,i\rpar\,S_{n+i}}{\sum_{i=i_0}^k\,W^t\lpar n,k,i\rpar},
$ where
$
W^t(n,k,i) = (-1)^i\,
\binom{k}{i}
\frac{\lpar n+i\rpar_{k-1}}{\Delta S_{n+i-1}}.
$
%--
\item Generalized Levin $y\,$-transform~\cite{Sidi}:
%--
%\bq
%y^n_k = \frac{\sum_{i=i_0}^k\,W^{y}\lpar n,k,i\rpar\,S_{n+i}}{\sum_{i=i_0}^k\,
%W^{y}\lpar n,k,i\rpar},
%\eq
%%--
%\[
%W^{y}(n,k,i) = (-1)^i\,
%\left(
%\begin{array}{c}
%k \\
%i
%\end{array}
%\right)
%\frac{\lpar n+i\rpar_{k-2}}{\Delta S_{n+i-1}}.
%\]
%
$
y^n_k = \frac{\sum_{i=i_0}^k\,W^{y}\lpar n,k,i\rpar\,S_{n+i}}{\sum_{i=i_0}^k\,
W^{y}\lpar n,k,i\rpar},
$ with
$
W^{y}(n,k,i) = (-1)^i\,
\binom{k}{i}
\frac{\lpar n+i\rpar_{k-2}}{\Delta S_{n+i-1}}.
$%\]
%--
\eei
%--
\subsection{Example applications of sequence transformations}
\label{sec:exampleseqtrans}

To discuss our results, we introduce the following notation:
%--
$
S_{\ssN,n} = \sum_{k=0}^n\,\ugm_k\,z^k + \sum_{k=n+1}^{\ssN}\,\ougm_k\,z^k
$,
%--
and $\tau_{\ssN,n}$ constructed accordingly.
For example, $\tau_{6,3} = N_{6,3}/D_{6,3}$ with
%--
\bqa
N_{6,3} &=& - \frac{720}{S_1 - S_0}\,S_1 + \frac{10\,800}{S_2 - S_1}\,S_2
            - \frac{50\,400}{S_3 - S_2}\,S_3 + \frac{100\,800}{S_{4,3} - S_3}\,S_{4,3}
%\nl
%{}&-&      
            -   \frac{90\,720}{S_{5,3} - S_{4,3}}\,S_{5,3}
            + \frac{30\,240}{S_{6,3} - S_{5,3}}\,S_{6,3},
\nl
D_{6,3} &=& - \frac{720}{S_1 - S_0}\,S_1 + \frac{10\,800}{S_2 - S_1}
            - \frac{50\,400}{S_3 - S_2}\,S_3 + \frac{100\,800}{S_{4,3} - S_3}
%\nl
%{}&-&         
            - \frac{90\,720}{S_{5,3} - S_{4,3}}
            + \frac{30\,240}{S_{6,3} - S_{5,3}}.
\eqa
%--

The transformations listed in \Sref{sec:otherseqtrans} have been applied to a suite of test series.
Note that for the calculations one could have used readily-available software, \eg the one described in \Bref{FFS},
Maple~\cite{maple}, or GSL~\cite{gsl}.
The suite of test series considered includes:

%--
\bei
%--
\item We first considered the series
%--
\bq 
S_{\infty}= \lpar 1 + z\rpar^{\nu} = 1 + \sum_{n=1}^{\infty}\ugm_n\,z^n, \qquad \nu= 12.62
\label{mim}
\eq
%--
where $\nu$ was tuned such that its first $3$ coefficients are similar to those of the series in \eqn{FS}.
%--
\begin{table}[h]
\[
\begin{array}{ccc|ccc}
\hline
\ugm_1 & \ugm_2 & \ugm_3 & \ugm_4 & \ugm_5 & \ugm_6 \\
12.620   & 73.322   & 259.56   & 624.24   & 1\,076.2   & 1\,366.8   \\
\cline{4-6}
         &          &          & {\overline{\ugm_4}} & {\overline{\ugm_5}} & {\overline{\ugm_6}} \\
         &          &          & 624.89   & 1\,081.8   & 1\,388.9 \\
\hline
\end{array}
\]
\caption[]{Actual and predicted coefficients for the series of \eqn{mim}, which was designed so as to approximately reproduce the values in \Tref{tab:ggFinputs} when $z=0.1$.}
\label{tab:sinfgamma}
\end{table}
%--
The sum of the series for $z= 0.1$ is $S_{\infty} = 3.32947445$. Using up to 
$6$ $\ugm\,$-coefficients, shown in \Tref{tab:sinfgamma}, we derive that the best improvement for the rate of convergence is obtained with the Levin $\tau\,$-transform of \eqn{Levintau} with $\beta=n=0$:
%--
\bq
S_{\infty} = 3.329\,474\,45,
\qquad
S_6        = 3.329\,335\,63,
\qquad
\tau_6   = 3.329\,474\,45 
\eq
%--
\Tref{tab:sinfgamma} also shows the partial results of using our recursive approximations algorithm.
Eventually, we obtain ${\overline\tau}_6 = 3.329\,622\,98$, or ${\overline\tau}_6/S_{\infty}-1=0.0045\%$.

\item The goodness of the approximation has also been tested by expanding the hypergeometric function ${}_2F_1( n+1/2,n+1;n+3/2;z^2)$ for large values of $n$, with positive results: in all cases convergence is improved.
%--
\item Several other examples, $\lpar 1 + z\rpar^{1/2}$, $\ln (1 + z)$, $e^z$, $\sum_{n=0}^{\infty}\,(-1)^n\,n\,!\;z^n$, $\Phi\lpar n,z,a\rpar$, where $\Phi$ is the Learch Phi-function, can be found in \Bref{Roy:2006cb}.
The same work provides examples where higher-order coefficients are estimated, \eg, $a_{\PGm,\Pe}$ (muon or electron $g-2$) and the hadronic ratio $R$.

\item Consider now the case of an asymptotic series, \eg
%--
\bq
S_{\infty}= \sum_{n=0}^{\infty}\,n\,!\;z^{n+1} = e^{-1/z}\,\mathrm{Ei}\lpar 1/z \rpar
\label{ei}
\eq
%--
where the exponential integral is a single-valued function in the plane cut along the negative real axis.
However, for $z > 0$, $\mathrm{Ei}(z)$ can be computed to great accuracy using several Chebyshev expansions.
Note that the {r.h.s.} of \eqn{ei} is the Borel sum of the series. 

The approximation returned by the $\ougm_n$ is not of high quality.
Nevertheless, the approximation works reasonably well and $\tau_{6,3}$ is not worse that $S_{6,3}$, as shown in \Tref{tab:ei}.
%--
\begin{table}[h]
\[
\begin{array}{c|cc|cc}
\hline
S_{\infty} & S_6        & \tau_6     & S_{6,3}    & \tau_{6,3} \\
\hline
1.097\,737\,72 & 1.097\,437\,00 & 1.097\,788\,64 & 1.097\,059\,09 & 1.097\,052\,34 \\
    {-}    & 0.027\%    & 0.005\%    & 0.062\%    & 0.062\%     \\
\hline
\end{array}
\]
\caption[]{Predictions for the series of \eqn{ei}.}
\label{tab:ei}
\end{table}
%--
It has been shown in \Bref{Sidi} that there is a large class of series that have Borel sums that are analytic in the cut-plane and the numerical results of \Bref{SF} suggest that  Levin-Weniger transforms produce approximations to these Borel sums.
Furthermore, in \Bref{Jentschura:1999vn}, numerical evidence is shown suggesting that the Weniger transform can resum a function with singularities in the Borel plane (but not on the positive axis).

\item Other relevant examples are: the prediction for the fifth (known) coefficient of the $\beta\,$-function of the Higgs boson coupling, the derivative expansion of QED effective action, and the partition function for zero-dimensional $\phi^4$ theory~\cite{Jentschura:1999vn}. 

\item We have also tested the method against some recent calculations like the leptonic contributions to the effective electromagnetic coupling at four-loop order in QED. The coefficients of $\alpha/\pi$ and $\Delta\alpha_{\mathrm lep}$ are~\cite{Sturm:2013uka}: $\ugm_1= 13.526\,31(8)$, $\ugm_2= 14.385\,53(6)$, and $\ugm_3= 84.828\,5(7)$.
The predicted and known results for $\ugm_4$ are $\ougm_4 = 705.22$ and $\ugm_4 = 770.76$, for a relative difference of $\ougm_4/\ugm_4 -1 = -8.5\%$.

\item There are cases where the algorithm cannot make a reliable prediction, such as in predicting QCD corrections to the QED $\beta\,$-functions, see \Brefs{Baikov:2012zm,Kataev:2012rf}.
Looking at Eqs.($4.4-4.6$) of \Bref{Baikov:2012zm} we see series with sudden jumps of sign in the coefficients; for instance, the series for $5$ flavors is
%--
\bq
\alpha^2\,\lpar 1.667 + 1.667\,a_{\ssS} + 2.813\,a^2_{\ssS} - 5.971\,a^3_{\ssS} -
32.336\,a^4_{\ssS}\rpar
\label{QEDbeta}
\eq
%--
with $a_{\ssS}= \alphas/\pi$. Our results, based on $\tau_{4,3}$ are shown in \Tref{tab:QEDbeta}.
%--
\begin{table}[h]
\[
\begin{array}{c|cc}
\hline
n_f & S_4 & \tau_{4,3} \\
\cline{1-3}
4   & 1.145\,469 & 1.146\,096 \\
5   & 1.138\,618 & 1.140\,940 \\
\cline{1-3}
\end{array}
\]
\caption[]{Predictions for the series of \eqn{QEDbeta}.}
\label{tab:QEDbeta}
\end{table}
%--
Here, neglecting the term ${\cal O}(a^4_{\ssS})$ or computing $\tau_4$ with an approximated $\ougm_4$ gives a difference of the same size.
In this case we are considering a series representing a self-energy that will have a two-particle cut (with the corresponding series of corrections), a three-particle cut 
(with the corresponding series of corrections), etc.
Therefore, at each order in perturbation theory new contributions (\ie new series) will arise and it is unsafe to make a guess by using only the first $3\,$ orders.
However, in this case, using $\ougm_4$ as an estimate of the uncertainty in $S_3$ gives a reasonable result:
%--
$%\bq
0.95 < |\tau_4 - S_4|/\ugm_4\,a^4_{\ssS} < 1.14.
$%\eq

%--
\eei

Further examples of the performance of different transforms on a number of test sequences can be found in \Brefs{Roy2,Roy3,Roy4,Roy5,WC,WCV,WNA}.

%Afer having studied all the examples above, we have retained two transforms when discussing MHOU: the Levin $\tau\,$-transform \eqn{Levintau} and the Weniger $\delta\,$-transform \eqn{Wenigerdelta}.

%% file: ggF.tex
%--
\section{Application to $\Pg\Pg\,$-fusion \label{App}}
\label{sec:ggF}
%--

For all the examples considered in \Sref{sec:exampleseqtrans} we have found that the Levin $\tau\,$-transform and the Weniger $\delta\,$-transform provide the fastest convergence.
The power of these transformations is due to the fact that the explicit estimates for the truncation error of the series are incorporated into the convergence acceleration. 
The Levin $\tau\,$-transform has been shown to work with good accuracy for the prediction of higher order coefficients of alternating and non-alternating factorially-divergent perturbation series, see \Bref{Jentschura:1999vn}.

Arguments supporting the general applicability of Levin transforms to the series of mathematical structures expected from quantum field theory can be found also in \Bref{Jentschura:1999vn}.

It should be noted that in the $\tau^n_k$ sequence transformation, the superscript $n$ indicates the minimal index occurring in the finite subset of input data, while $k$, the order of the transformation, is a measure of the complexity for the transformation itself.
It is worth noting that $\tau^n_k$ requires knowledge of the first $n+k$ partial sums, that is why we limit our considerations to $\tau_k \equiv \tau^0_k$.

The most important question concerns the reliability of the procedure when applied
to the series of \eqn{FS}.
%--
\subsection{Applicability}
In motivating the applicability of the procedure, scale variation can be of use.
Consider
%--
\bq
\sigma^n_{\Pg\Pg}\lpar \tau,\mu\rpar = \sigma^0_{\Pg\Pg}\lpar \tau,\mu\rpar\,
S_{n,3}\lpar \mu\rpar,
\quad 
S_{n,3}\lpar \mu\rpar = 1 + \sum_{k=1}^3\,\alphas^k(\mu)\,\ugm_k(\mu) + \sum_{k=4}^n\,\alphas^k(\mu)\,\ougm_k(\mu)
\eq
%--
where $\tau= M^2_{\PH}/s$, $\sqrt{s} = 8\UTeV$, and vary the QCD scales with $\xi = 2$. 
Introducing
%--
\bq
\sigma^{-,n}_{\xi} = \min\{
\sigma^n\lpar Q,\mu/\xi\rpar\,,\,
\sigma^n\lpar Q,\xi\,\mu\rpar\},
\quad
\sigma^{+,n}_{\xi} = \max\{
\sigma^n\lpar Q,\mu/\xi\rpar\,,\,
\sigma^n\lpar Q,\xi\,\mu\rpar\}
\label{defspm}
\eq
%--
and $D_n=1-\sigma^{-,n}_2/\sigma^{+,n}_2$
we obtain the values reported in \Tref{QCDsv}.
Comparing the results for $D_2$ and $D_3$ it can be seen that the variability due to scale variation is substantially reduced by the inclusion of the N${}^3$LO term.
We expect that a reliable estimate of the missing higher orders should follow the trend of further reducing the effect as is the case. 
%--
\begin{table}
\[
\begin{array}{c|ccc}
\hline
D_n  &  \ugm^{\cent}_3 - \Delta\ugm_3 & \ugm^{\cent}_3  & \ugm^{\cent}_3 + \Delta\ugm_3 \\ 
\hline
D_2  &   & 27.02\,\% &  \\
D_3  &  14.93\,\% & 16.08\,\% & 17.21\,\% \\
\hline
D_4  &   6.59\,\% &  7.99\,\% &  9.68\,\% \\
D_5  &   2.20\,\% &  3.08\,\% &  4.59\,\% \\
D_6  &   0.14\,\% &  0.39\,\% &  1.38\,\% \\
\hline
\end{array}
\]
\caption{Effect of QCD scale variation for predicted higher order terms in the Higgs gluon-gluon fusion production cross-section.
$D_n=1-\sigma^{-,n}_2/\sigma^{+,n}_2$ with the $\sigma^{\pm,n}_\xi$ are defined in \eqn{defspm}.
In the extrapolation region ($n\geq3$) the variation decreases as expected from a reliable estimate of MHO.
}
\label{QCDsv}
\end{table}
%--
The coefficients of the perturbative series, computed with $\tau_k$ at $\mu= \mh$, are given 
in \Tref{xxx}.
%--
\begin{table}
\[
\begin{array}{c|ccc|ccc}
\hline
\ougm_n & & \mathrm{Levin-}\tau &  & & \mathrm{Weniger-}\delta &  \\
 &  \ugm^{\cent}_3 - \Delta\ugm_3 & \ugm^{\cent}_3  & \ugm^{\cent}_3 + \Delta\ugm_3 &  \ugm^{\cent}_3 - \Delta\ugm_3 & \ugm^{\cent}_3  & \ugm^{\cent}_3 + \Delta\ugm_3 \\ 
\hline
\ougm_4 & 1\,437.9 & 1\,806.6 & 2\,214.7 & 1\,512.2 & 1\,860.8 & 2\,244.3\\
\ougm_5 & 5\,412.4 & 8\,185.6 & 11\,733.0 & 6\,276.6 & 8\,912.3 & 12\,183.0 \\
\ougm_6 & 18\,979.0 & 35\,677.0 & 61\,133.0 & 25\,243.0 & 41\,918.0 & 65\,605.0 \\
\hline
\end{array}
\]
\caption[]{Predicted higher-order coefficients in gluon-gluon--fusion, computed 
at $\mu= \mh$.
}
\label{xxx}
\end{table}
%--
The ratio $R_n = \alphas\,\ougm_{n+1}/\ougm_n$ becomes constant to a very good approximation,
and is given in \Tref{Kasy}, where $\delta_n$ is defined in \eqn{Wenigerdelta}.
%--
\begin{table}
\[
%\begin{array}{c|cc}
%\hline
%R_n & \mathrm{Levin\ } \tau & \mathrm{Weniger\ } \delta \\        
%\hline
%R_0 & 1.3280 & 1.3280 \\
%R_1 & 0.6800 & 0.6800 \\
%R_2 & 0.5836 & 0.5360 \\
%\hline
%R_3 & 0.5354 & 0.4880 \\
%R_4 & 0.5065 & 0.4640 \\
%R_5 & 0.4873 & 0.4496 \\
%\hline
%\end{array}
\begin{array}{c|ccc|ccc}
\hline
R_n & R_0 & R_1 & R_2 & R_3 & R_4 & R_5 \\
\hline
\mathrm{Levin-}\tau     & 1.3280 & 0.6800 & 0.5836 & 0.5354 & 0.5065 & 0.4873\\
\mathrm{Weniger-}\delta & 1.3280 & 0.6800 & 0.5360 & 0.4880 & 0.4640 & 0.4496 \\
\hline
\end{array}
\]
\caption[]{$R_n = \alphas\,\ougm_{n+1}/\ougm_n(\ugm_n)$ for the $\tau$ and $\delta$ 
transforms (note that the denominator is not an extrapolation when available). 
It can be seen that $R_n$ is constant to better than 
10\% in the extrapolation region ($n\geq3$) for both transforms.
}
\label{Kasy}
\end{table}

Note that this does not represent a formal proof that there is an upper bound on the remainder
but makes plausible the argument in favor of that.
%--
\subsection{Numerical results}
\label{sec:ggFcalc}
%--
Our strategy for estimating MHO and MHOU can be summarized as follows: we select a scale, 
$\mu = \mh$, for $\Pg\Pg\,$-fusion, and estimate the uncertainty due to higher orders 
at that scale.
This implies that the (scale variation) uncertainty at the chosen scale is part of the 
uncertainty due to higher orders and should not be counted twice.
Therefore, we compare
%--
\bq
\sigma^{X,n}_{\Pg\Pg} = \sigma^0_{\Pg\Pg}\lpar \mu = \mh\rpar\,
X_{n,3}\lpar \mu = \mh\rpar,
\qquad\mbox{with}\qquad
X\in\{\ssS, \tau, \delta\}
\label{sigapp}
\eq
%--
and report the result of the calculations in \Tref{Tapp} where the coefficients needed to 
construct $\sigma^{\ssS,n}_{\Pg\Pg}$ are based on $\tau\,$-transform.
To understand the comparison one should bear in mind that sequence transforms can also
be characterized by the highest coefficient involved: $\tau_k$ requires $\ougm_k$ but
$\delta_k$ requires $\ougm_{k+1}$.
Therefore, we expect $\tau_k$ and $\delta_{k-1}$ to give predictions with comparable quality.
%--
\begin{table}
\[
\begin{array}{c|ccc|ccc|ccc}
\hline
n & & \sigma^{\ssS,n\;}_{\Pg\Pg}\,[\Upb] & & & \sigma^{\tau,n\;}_{\Pg\Pg}\,[\Upb] & & & 
\sigma^{\delta,n-1}_{\Pg\Pg}\,[\Upb] & \\
 & \ugm^\cent_3-\Delta\ugm_3 & \ugm^\cent_3 & \ugm^\cent_3+\Delta\ugm_3 & \ugm^\cent_3-\Delta\ugm_3 & \ugm^\cent_3 & \ugm^\cent_3+\Delta\ugm_3 & \ugm^\cent_3-\Delta\ugm_3 & \ugm^\cent_3 & \ugm^\cent_3+\Delta\ugm_3  \\
\hline
3 & 19.89889 & 20.12922 & 20.35954 & 21.83017 & 23.07444 & 24.83209 & & & \\
4 & 21.10181 & 21.64063 & 22.21236 & 21.92044 & 23.10458 & 24.79751 & 22.21881 & 23.42508 & 25.07780\\
5 & 21.60801 & 22.40620 & 23.30967 & 21.91988 & 23.10473 & 24.79661 & 22.21864 & 23.42407 & 25.07535\\
6 & 21.80644 & 22.77922 & 23.94883 & 21.91988 & 23.10473 & 24.79658 & 22.21864 & 23.42405 & 25.07525\\    
\hline
\end{array}
\]
\caption[]{Cross-sections obtained using \eqn{sigapp}, using $\mu=\mh$. For $\sigma^{\delta,n}_{\Pg\Pg}$, $\beta=1$ is used for the Weniger $\delta\,$-transform. Note that in the case of the Weniger transform the index is shifted so that rows represent the same order in $\ougm_n$. }
\label{Tapp}
\end{table}
%--
The results show that using the Levin $\tau\,$-transform improves the convergence; indeed
$n = 3$ is already a good approximation with $\sigma^{\tau,6}_{\Pg\Pg}/\sigma^{\tau,3}_{\Pg\Pg}-1=0.13\%$. 
The use of other transforms is compatible with $\tau_6$ to within 2\%: if we use the Weniger 
$\delta\,$-transform of \eqn{Wenigerdelta} (with $\beta = 1$) we obtain
$\sigma^{\delta,5}_{\Pg\Pg}/\sigma^{\tau,6}_{\Pg\Pg} - 1 = 1.38\%$.

Additionally, we have investigated the use of $\beta\,$-tuning, using the Levin $\tau\,$-transform \eqn{Levintau} with $\beta\neq0$.
To have $\ugm_3 = \ugm^{\cent}_3$, we find $\beta = -0.2482$, and calculate $\sigma^{\tau,5}_{\Pg\Pg}(\beta) = 23.542\Upb$.
This is to be compared with $\sigma^{\tau,5}_{\Pg\Pg}(\beta=0) = 23.105\Upb$, the difference being within the uncertainty induced by $\Delta\ugm_3$. 
Our conclusion is that $\beta\,$-tuning is a procedure to be adopted in those cases where there is a reasonable guess on the value of the next coefficient or on the interval where it is expected.
Furthermore, all cases where the $\beta\,$-tuned results are substantially different 
from $\beta = 0$ should be taken with the due caution.
Finally, basing the whole procedure on $\delta\,$-transforms or estimating the coefficients with 
$\delta_k(1)$ and accelerating the series with $\tau_6$ gives consistent results, namely 
$23.4241\Upb$ (with $\delta_5$) in the first case and $23.4253\Upb$ in the second.  

It is worth noting that if any of the transforms predicts at least one extra coefficient of the series, then in principle the whole function is known, which is unlikely to be the case in any physical problem.
We can only conclude that a judicious use can make predictions at some relatively good level of accuracy.
We also know that all transforms basically differ in the choice of the remainder estimates.
A good choice should satisfy the following asymptotic condition~\cite{Weniger:2001ct}:
%--
$%\bq
R_n = \frac{S_{\infty} - S_n}{\omega_n} \sim c$, when $n\to \infty
$%\eq
%--
, where $\omega_n$ is the remainder.
Levin selects $\omega_n = \Delta S_{n-1}$ and, from \refT{Kasy}, we derive an approximate relation
%--
$%\bq
\ugm_{n+1}\,\alphas \approx K\,\ugm_n$, for $n > n_0
$%\eq
%--
, where $K$ is a constant with $K < 1$.
In this case $R_n \to 1/(1-K)$ for sufficiently large $n$.
%--
\subsection{Discussion of MHOU}
\label{sec:discussionMHOU}

Given that the sequence transform procedures outlined above provide an estimate for the sum of the full series, when estimating the uncertainty on that quantity we will be deliberately conservative.

\paragraph{Uncertainty due to MHO estimation}
Given the different nature of the calculations represented by $\sigma^{\ssS,3}_{\Pg\Pg}$ and $\sigma^{\delta,5}_{\Pg\Pg}$, it can be expected that, to a very good accuracy, $\sigma^{\ssS,3}_{\Pg\Pg}<\sigma_{\Pg\Pg}<\sigma^{\delta,5}_{\Pg\Pg}$.
For $\ugm_3=\ugm^\cent_3$, this defines the interval $[20.13\,,\,23.42]\Upb$ that has a relative width of $16.4\%$. 
For comparison, the N${}^3$LO calculation for $\mu= \mh$ and $\ugm_3=\ugm^\cent_3$ yields $\sigma_{\Pg\Pg} = 20.13\Upb$ and traditional QCD scale variations with $\xi = 2$ leads to the interval $[18.90\,,\,21.93]\Upb$ that has a relative width of $16.1\%$ (the authors of \Bref{Ball:2013bra} quote $\pm 7\%$).
It is worth noting how in our approach the interval is shifted by $\approx\,+7\%$ with respect to the N${}^3$LO result.
This is to be compared to the $+17\%$ of N${}^3$LO with respect to NNLO~\cite{Ball:2013bra}.
%
%For comparison, QCD scale variation
%gives $-7.1\,+7.8\%$~\cite{Dittmaier:2011ti}.

\paragraph{Uncertainty due to $\Delta\ugm_3$}
We can now discuss how to take into account the uncertainty on $\ugm_3$ induced by the $\Delta\ugm_3\lpar \mu = \mh\rpar$.
In line with a simple and conservative approach that can later be refined, we consider all values of $\ugm_3$ in the interval $[\ugm^\cent_3-\Delta\ugm_3,\ugm^\cent_3+\Delta\ugm_3]$ as equally likely and take the lowest value of $\sigma^{\ssS,3}_{\Pg\Pg}$ and the highest value of $\sigma^{\delta,5}_{\Pg\Pg}$.

\paragraph{Result}
The previous choices lead to an interval with a relative width of $26.01\%$, shifted by at least $+5\%$ with respect to the N${}^3$LO result:
\bq
\sigma_{\Pg\Pg} \in \left[ \sigma^{\ssS,3}_{\Pg\Pg}(\ugm^\cent_3-\Delta\ugm_3)\,,\,
\sigma^{\delta,5}_{\Pg\Pg}(\ugm^\cent_3+\Delta\ugm_3)\right] = 
\left[ \sigma_- \,,\, \sigma_+\right] = 
\left[ 19.89889\,,\,25.07525\right] \Upb.
\label{bg}
\eq

To conclude, our prediction is that the ``true'' cross-section value is bracketed by the estimations of \eqn{bg} as all other transforms fall in that interval.
For instance, $J^2_1$ from \eqn{bre} gives $23.018\Upb$ and $\tau^1_4$ gives $23.244\Upb$.

%The advantages of the method 
The advantages of our recipe for estimating MHOU are that the result does not depend on the choice of the parameter expansion (it is based on partial sums) and it takes into account the nature of the coefficients, \ie, that the known terms of the perturbative expansion in $\Pg\Pg\,$-fusion are positive.
Starting from the proposal in \eqn{bg}, the corresponding pdf can be derived following the work of \Bref{Cacciari:2011ze}.

%The MHOU proposal of \eqn{bg} deliberately treats the effect of different MHO estimations and of $\Delta\ugm_3$ in a simple way.
%We are aware that more refined uses of these inputs can be made, \eg, by parameterizing the effect of MHO using $\beta$, which is fundamentally related to the proposed MHO estimation procedure.
%The choice of pdf in $\beta$-space, such as flat for different MHO estimations, can then be carried to cross-section--space, where the resulting pdf would not necessarily be flat.

%\CheckG{
%The corresponding pdf could be derived by following the work of \Bref{Cacciari:2011ze} giving
%%--
%\[
%P_{\rm CH}(\sigma) = N^{-1}_{\sigma}\,\left\{
%\begin{array}{ll}
%\lpar \frac{\Delta\sigma}{\sigma_+ - \sigma}\rpar^5 & \mbox{if}\qquad \sigma < \sigma_- \\
%1                                        & \mbox{if}\qquad \sigma_- < \sigma < \sigma_+ \\
%\lpar \frac{\Delta\sigma}{\sigma - \sigma_-}\rpar^5 & \mbox{if}\qquad \sigma > \sigma_+ \\
%\end{array}
%\label{opt0}
%\right.
%\]
%%--
%where 
%$\Delta\sigma= \sigma_+ - \sigma_-$ and $N_{\sigma} = \frac{3}{2}\,\Delta{\sigma}$. 
%%--
%}

%% file: main.bbl
\begin{thebibliography}{10}
\expandafter\ifx\csname url\endcsname\relax
  \def\url#1{\texttt{#1}}\fi
\expandafter\ifx\csname urlprefix\endcsname\relax\def\urlprefix{URL }\fi
\expandafter\ifx\csname href\endcsname\relax
  \def\href#1#2{#2} \def\path#1{#1}\fi

\bibitem{Dittmaier:2011ti}
S.~Dittmaier, et~al., {Handbook of LHC Higgs Cross Sections: 1. Inclusive
  Observables}\href {http://arxiv.org/abs/1101.0593} {\path{arXiv:1101.0593}}.

\bibitem{Cacciari:2011ze}
M.~Cacciari, N.~Houdeau, {Meaningful characterisation of perturbative
  theoretical uncertainties}, JHEP 1109 (2011) 039.
\newblock \href {http://arxiv.org/abs/1105.5152} {\path{arXiv:1105.5152}},
  \href {http://dx.doi.org/10.1007/JHEP09(2011)039}
  {\path{doi:10.1007/JHEP09(2011)039}}.

\bibitem{THUTFindico}
{THUTF meeting: Missing higher orders and PDF uncertainties},
  \url{https://indico.cern.ch/conferenceDisplay.py?confId=251810}, [Online;
  accessed 3-July-2013] (2013).

\bibitem{mhovertwo}
C.~Anastasiou, K.~Melnikov, {Higgs boson production at hadron colliders in NNLO
  QCD}, Nucl.Phys. B646 (2002) 220--256.
\newblock \href {http://arxiv.org/abs/hep-ph/0207004}
  {\path{arXiv:hep-ph/0207004}}, \href
  {http://dx.doi.org/10.1016/S0550-3213(02)00837-4}
  {\path{doi:10.1016/S0550-3213(02)00837-4}}.

\bibitem{softgluonresummation}
S.~Catani, D.~de~Florian, M.~Grazzini, P.~Nason, {Soft gluon resummation for
  Higgs boson production at hadron colliders}, JHEP 0307 (2003) 028.
\newblock \href {http://arxiv.org/abs/hep-ph/0306211}
  {\path{arXiv:hep-ph/0306211}}.

\bibitem{Simon:1972yi}
B.~Simon, {Summability methods, the strong asymptotic condition, and unitarity
  in quantum field theory}, Phys. Rev. Lett. 28 (1972) 1145--1146.
\newblock \href {http://dx.doi.org/10.1103/PhysRevLett.28.1145}
  {\path{doi:10.1103/PhysRevLett.28.1145}}.

\bibitem{Bender:1971gu}
C.~M. Bender, T.~T. WU, {Large order behavior of Perturbation theory}, Phys.
  Rev. Lett. 27 (1971) 461.
\newblock \href {http://dx.doi.org/10.1103/PhysRevLett.27.461}
  {\path{doi:10.1103/PhysRevLett.27.461}}.

\bibitem{Bender:1990pd}
C.~M. Bender, T.~Wu, {Anharmonic oscillator. 2: A Study of perturbation theory
  in large order}, Phys.Rev. D7 (1973) 1620--1636.
\newblock \href {http://dx.doi.org/10.1103/PhysRevD.7.1620}
  {\path{doi:10.1103/PhysRevD.7.1620}}.

\bibitem{Lipatov:1977hj}
L.~Lipatov, {Divergence of Perturbation Series and Pseudoparticles}, JETP Lett.
  25 (1977) 104--107.

\bibitem{Buchvostov:1977fi}
A.~Bukhvostov, L.~Lipatov, {High Orders of the Perturbation Theory in Scalar
  Electrodynamics}, Phys.Lett. B70 (1977) 48--50.
\newblock \href {http://dx.doi.org/10.1016/0370-2693(77)90341-0}
  {\path{doi:10.1016/0370-2693(77)90341-0}}.

\bibitem{Brezin:1976wa}
E.~Brezin, J.-C. Le~Guillou, J.~Zinn-Justin, {Perturbation Theory at Large
  Order. 2. Role of the Vacuum Instability}, Phys.Rev. D15 (1977) 1558--1564.
\newblock \href {http://dx.doi.org/10.1103/PhysRevD.15.1558}
  {\path{doi:10.1103/PhysRevD.15.1558}}.

\bibitem{Brezin:1976vw}
E.~Brezin, J.~Le~Guillou, J.~Zinn-Justin, {Perturbation Theory at Large Order.
  1. The $\varphi^{2N}$ Interaction}, Phys.Rev. D15 (1977) 1544--1557.
\newblock \href {http://dx.doi.org/10.1103/PhysRevD.15.1544}
  {\path{doi:10.1103/PhysRevD.15.1544}}.

\bibitem{Vainshtein:1994qq}
A.~Vainshtein, V.~I. Zakharov, {Ultraviolet renormalon reexamined}, Phys. Rev.
  D54 (1996) 4039--4048.
\newblock \href {http://dx.doi.org/10.1103/PhysRevD.54.4039}
  {\path{doi:10.1103/PhysRevD.54.4039}}.

\bibitem{Suslov:2005zi}
I.~Suslov, {Divergent perturbation series}, Zh.Eksp.Teor.Fiz. 127 (2005) 1350.
\newblock \href {http://arxiv.org/abs/hep-ph/0510142}
  {\path{arXiv:hep-ph/0510142}}, \href {http://dx.doi.org/10.1134/1.1995802}
  {\path{doi:10.1134/1.1995802}}.

\bibitem{Fischer:1995cx}
J.~Fischer, {High order behavior and summation methods in perturbative QCD},
  Acta Phys. Polon. B27 (1996) 2549--2566.
\newblock \href {http://arxiv.org/abs/hep-ph/9512269}
  {\path{arXiv:hep-ph/9512269}}.

\bibitem{Hardy}
G.~H. Hardy, {Divergent Series}, Oxford: Clarendon Press. (1949) {}ISBN
  978-0-8218-2649-2.

\bibitem{Zakharov:1992bx}
V.~I. Zakharov, {QCD perturbative expansions in large orders}, Nucl. Phys. B385
  (1992) 452--480.
\newblock \href {http://dx.doi.org/10.1016/0550-3213(92)90054-F}
  {\path{doi:10.1016/0550-3213(92)90054-F}}.

\bibitem{Sokal:1980ey}
A.~Sokal, {An Improvement of Watson's Theorem on Borel Summability}, J. Math.
  Phys. 21 (1980) 261--263.
\newblock \href {http://dx.doi.org/10.1063/1.524408}
  {\path{doi:10.1063/1.524408}}.

\bibitem{Beneke:1992ea}
M.~Beneke, V.~I. Zakharov, {Improving large order perturbative expansions in
  quantum chromodynamics}, Phys. Rev. Lett. 69 (1992) 2472--2474.
\newblock \href {http://dx.doi.org/10.1103/PhysRevLett.69.2472}
  {\path{doi:10.1103/PhysRevLett.69.2472}}.

\bibitem{Collins:1977dw}
J.~C. Collins, D.~E. Soper, {Large Order Expansion in Perturbation Theory},
  Annals Phys. 112 (1978) 209--234.
\newblock \href {http://dx.doi.org/10.1016/0003-4916(78)90084-2}
  {\path{doi:10.1016/0003-4916(78)90084-2}}.

\bibitem{CVZ}
H.~Cohen, F.~Villegas, D.~Zagier, {Convergence Acceleration of Alternating
  Series}, Experimental Mathematics 9:1 (2000) 3.

\bibitem{BZ}
C.~Brezinski, M.~Redivo~Zaglia, {Extrapolating Methods. Theory and Practice},
  North-Holland.

\bibitem{BF}
C.~Brezinski, {Acceleration des suites a convergence logarithmique}, C. R.
  Acad. Sci. Paris 273A (1971) 727--730.

\bibitem{SB}
A.~Sidi, {An algorithm for a special case of a generalization of the Richardson
  extrapolation process}, Numerische Mathematik 38 (1982) 299--307.

\bibitem{SBB}
A.~Sidi, {Borel Summability and Converging Factors for Some Everywhere
  Divergent Series}, SIAM J. Math. Anal. 17(5) (1986) 1222--1231.

\bibitem{Moch:2005ky}
S.~Moch, A.~Vogt, {Higher-order soft corrections to lepton pair and Higgs boson
  production}, Phys. Lett. B631 (2005) 48--57.
\newblock \href {http://arxiv.org/abs/hep-ph/0508265}
  {\path{arXiv:hep-ph/0508265}}, \href
  {http://dx.doi.org/10.1016/j.physletb.2005.09.061}
  {\path{doi:10.1016/j.physletb.2005.09.061}}.

\bibitem{Ball:2013bra}
R.~D. Ball, M.~Bonvini, S.~Forte, S.~Marzani, G.~Ridolfi, {Higgs production in
  gluon fusion beyond NNLO}\href {http://arxiv.org/abs/1303.3590}
  {\path{arXiv:1303.3590}}.

\bibitem{Buehler:2013fha}
S.~Buehler, A.~Lazopoulos, {Scale dependence and collinear subtraction terms
  for Higgs production in gluon fusion at N3LO}\href
  {http://arxiv.org/abs/1306.2223} {\path{arXiv:1306.2223}}.

\bibitem{Weniger:1997zz}
E.~J. Weniger, {Performance of superconvergent perturbation theory}, Phys. Rev.
  A56 (1997) 5165--5168.
\newblock \href {http://dx.doi.org/10.1103/PhysRevA.56.5165}
  {\path{doi:10.1103/PhysRevA.56.5165}}.

\bibitem{Jentschura:1999vn}
U.~Jentschura, J.~Becher, E.~Weniger, G.~Soff, {Resummation of QED perturbation
  series by sequence transformations and the prediction of perturbative
  coefficients}, Phys. Rev. Lett. 85 (2000) 2446--2449.
\newblock \href {http://arxiv.org/abs/hep-ph/9911265}
  {\path{arXiv:hep-ph/9911265}}, \href
  {http://dx.doi.org/10.1103/PhysRevLett.85.2446}
  {\path{doi:10.1103/PhysRevLett.85.2446}}.

\bibitem{Weniger:2001ct}
E.~J. Weniger, {Nonlinear sequence transformations: Computational tools for the
  acceleration of convergence and the summation of divergent series}\href
  {http://arxiv.org/abs/math/0107080} {\path{arXiv:math/0107080}}.

\bibitem{Roy:2006cb}
D.~Roy, R.~Bhattacharya, {Prediction of unknown terms of a sequence and its
  application to some physical problems}, Annals Phys. 321 (2006) 1483--1523.
\newblock \href {http://dx.doi.org/10.1016/j.aop.2005.12.010}
  {\path{doi:10.1016/j.aop.2005.12.010}}.

\bibitem{Zama}
Cizek, Zamastil, Sk\'ala, {New summation technique for rapidly divergent
  perturbation series. Hydrogen atom in magnetic field}, J. Math. Phys. 44
  (2003) 962--968.

\bibitem{Zamastil:2013jz}
J.~Zamastil, {Approximate recalculation of the $\alpha(\PZ \alpha)^5$
  contribution to the self-energy effect on hydrogenic states with a multipole
  expansion}, Annals Phys. 328 (2013) 139--157.
\newblock \href {http://dx.doi.org/10.1016/j.aop.2012.09.007}
  {\path{doi:10.1016/j.aop.2012.09.007}}.

\bibitem{Lev}
D.~Levin, {Development of non-linear transformations for improving convergence
  of sequences}, Int. J. Comput. Math. 3 (1973) 371--388.

\bibitem{Wen}
E.~J. Weniger, {Nonlinear sequence transformations for the acceleration of
  convergence and the summation of divergent series}, Comput. Phys. Rep. 10
  (1989) 189--371.

\bibitem{Wen2}
E.~J. Weniger, {Mathematical properties of a new Levin-type sequence
  transformation introduced by Cizek, Zamastil, and Skala. I. Algebraic
  theory}, J. Math. Phys. 45 (2004) 1209.

\bibitem{Wynn}
P.~Wynn, {On a device for computing the $e_m(S_n)$ transformation}, Math. Tab.
  Aids Comput. 10 (1956) 91--96.

\bibitem{Brez}
C.~Brezinski, {A general extrapolation algorithm}, Numer. Math. 35 (1980)
  175--187.

\bibitem{Sidi}
A.~Sidi, {A new method for deriving Pade approximants for some hypergeometric
  functions}, J. Comput. Appl. Math. 7 (1980) 37--40.

\bibitem{FFS}
T.~Fessler, W.~Ford, D.~Smith, {HURRY: An acceleration algorithm for scalar
  sequences and series}, ACM Trans. Math. Software 9 (1983) 346--354.

\bibitem{maple}
Maplesoft, {Maple Reference: Numerical Summation},
  \url{http://www.maplesoft.com/support/help/Maple/view.aspx?path=evalf/Sum},
  [Online; accessed 3-July-2013].

\bibitem{gsl}
{GNU Scientific Library Reference: Series Acceleration},
  \url{http://www.gnu.org/software/gsl/manual/html_node/Series-Acceleration.html},
  [Online; accessed 3-July-2013].

\bibitem{SF}
W.~Ford, D.~Smith, {Acceleration of linear and logarithmic convergence}, SIAM
  J. Num. Anal. 16 (1979) 223--240.

\bibitem{Sturm:2013uka}
C.~Sturm, {Leptonic contributions to the effective electromagnetic coupling at
  four-loop order in QED}\href {http://arxiv.org/abs/1305.0581}
  {\path{arXiv:1305.0581}}.

\bibitem{Baikov:2012zm}
P.~Baikov, K.~Chetyrkin, J.~Kuhn, J.~Rittinger, {Vector Correlator in Massless
  QCD at Order ${\mathcal O}(\alphas^4)$ and the QED beta-function at Five
  Loop}, JHEP 1207 (2012) 017.
\newblock \href {http://arxiv.org/abs/1206.1284} {\path{arXiv:1206.1284}},
  \href {http://dx.doi.org/10.1007/JHEP07(2012)017}
  {\path{doi:10.1007/JHEP07(2012)017}}.

\bibitem{Kataev:2012rf}
A.~Kataev, S.~Larin, {Analytical five-loop expressions for the renormalization
  group QED $\beta$-function in different renormalization schemes}, Pisma Zh.
  Eksp. Teor. Fiz. 96 (2012) 64--67.
\newblock \href {http://arxiv.org/abs/1205.2810} {\path{arXiv:1205.2810}},
  \href {http://dx.doi.org/10.1134/S0021364012130073}
  {\path{doi:10.1134/S0021364012130073}}.

\bibitem{Roy2}
D.~Roy, R.~Bhattacharya, S.~Bhowmick, {Rational approximants generated by the
  u-transform}, Comput. Phys. Commun. 78 (1993) 29--54.

\bibitem{Roy3}
D.~Roy, R.~Bhattacharya, S.~Bhowmick, {Rational approximants using
  Levin-Weniger transforms}, Comput. Phys. Commun. 93 (1996) 159--178.

\bibitem{Roy4}
D.~Roy, R.~Bhattacharya, S.~Bhowmick, {Rational interpolation using
  Levin-Weniger transforms}, Comput. Phys. Commun. 101 (1997) 213--222.

\bibitem{Roy5}
D.~Roy, R.~Bhattacharya, S.~Bhowmick, {Iterations of convergence accelerating
  nonlinear transforms}, Comput. Phys. Commun. 54 (1989) 31--46.

\bibitem{WC}
E.~Weniger, J.~Cizek, {Rational approximations for the modified Bessel function
  of the second kind}, Comput. Phys. Commun. 59 (1990) 471--493.

\bibitem{WCV}
E.~Weniger, J.~Cizek, F.~Vinette, {Very accurate summation for the infinite
  coupling limit of the perturbation series expansions of anharmonic
  oscillators}, Phys. Lett. A 156 (1991) 169--174.

\bibitem{WNA}
E.~Weniger, {Interpolation between sequence transformations}, Numer. Algor. 3
  (1992) 447--496.

\end{thebibliography}
